\documentclass[lettersize,journal]{IEEEtran}
\usepackage{amsmath,amsfonts}
\usepackage{algorithmic}
\usepackage{array}
\usepackage[caption=false,font=normalsize,labelfont=sf,textfont=sf]{subfig}
\usepackage{textcomp}
\usepackage{stfloats}
\usepackage{url}
\usepackage{verbatim}
\usepackage{graphicx}
\def\BibTeX{{\rm B\kern-.05em{\sc i\kern-.025em b}\kern-.08em
    T\kern-.1667em\lower.7ex\hbox{E}\kern-.125emX}}
\usepackage{balance}

\usepackage{arydshln}
\usepackage{algorithm}%
\usepackage{algorithmic}
\usepackage{xcolor}

\begin{document}
\title{Adaptive Decoding via Hierarchical Neural Information Gradients in Mouse Visual Tasks}
\author{Jingyi Feng, Xiang Feng
\thanks{Manuscript created October, 2025. The authors are at School of Computer Science, Wuhan University. E-mail: fjy2035@gmail.com (Jingyi Feng).}
}


\maketitle

\begin{abstract}
Understanding the encoding and decoding mechanisms of dynamic neural responses to different visual stimuli is an important topic in exploring how the brain represents visual information.
Currently, hierarchically deep neural networks (DNNs) have played a significant role as tools for mining the core features of complex data. However, most methods often overlook the dynamic generation process of neural data, such as hierarchical brain's visual data, within the brain's structure. 
In the decoding of brain's visual data, two main paradigms are 'fine-grained decoding tests' and 'rough-grained decoding tests', which we define as focusing on a single brain region and studying the overall structure across multiple brain regions, respectively. In this paper, we mainly use the Visual Coding Neuropixel dataset from the Allen Brain Institute, and the hierarchical information extracted from some single brain regions (i.e., fine-grained decoding tests) is provided to the proposed method for studying the adaptive topological decoding between brain regions, called the Adaptive Topological Vision Transformer, or AT-ViT. 
In numerous experiments, the results reveal the importance of the proposed method in hierarchical networks in the visual tasks, and also validate the hypothesis that "the hierarchical information content in brain regions of the visual system can be quantified by decoding outcomes to reflect an information hierarchy." Among them, we found that neural data collected in the hippocampus can have a random decoding performance, and this negative impact on performance still holds significant scientific value.
\end{abstract}

\begin{IEEEkeywords}
Adaptive decoding, fine-rough-grained test paradigm, hierarchical information gradients, random performance.
\end{IEEEkeywords}

\section{Introduction}

The encoding and decoding mechanisms of dynamic neural responses to different visual stimuli can promote the generation of intelligent behaviors similar to human beings in machines. In the field of neuroscience, the visual system is mainly used to receive a large amount of sensory input from the external world. After the complex organization of these inputs in different regions of the brain, it can participate in more high-level cognitive functions \cite{Decoding2024}. 
The transmission process of visual stimuli in the brain is roughly from the retina to the thalamus, then to the primary visual cortex \cite{Receptive1962}, and finally through two pathways to the anterior prefrontal nucleus (APN) \cite{Projections1995,Heterogeneous2007} and the hippocampus \cite{the2019}. Therefore, clarifying how visual information is encoded and decoded by a hierarchical structure has an important impact on understanding the computational principles of the visual system. 

\begin{figure}
    \centering
    \includegraphics[width=0.5\textwidth]{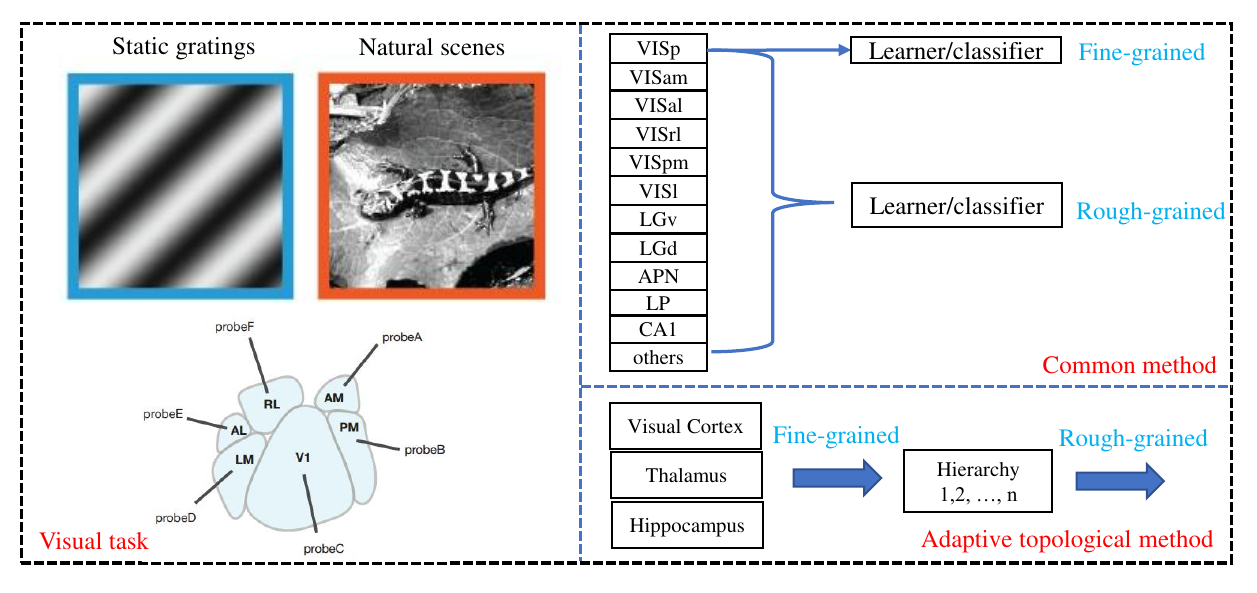}  
    \caption{To understand the neural processing of visual stimuli, various visual stimuli, including natural and artificial images/videos, can be presented to animals, and the respective neural responses from primary, secondary, and higher-order visual areas can be recorded.}
    \label{fig: intro}
\end{figure}

Visual stimuli are mainly transmitted in the brain in the form of neural impulses or electrical signals. 
In encoding, the neural responses of the brain to external stimulus, such as natural scenes and videos \cite{Retinal2022,Decoding2024}, etc., are mainly studied from a neurobiological perspective. In decoding, models are established through technical means to solve object classification tasks \cite{Neural2018} and pixel-level image reconstruction tasks \cite{Reconstruction2020}, etc. In existing visual decoding schemes, there are mainly studies on decoding neural data collected from a single brain region, such as the primary visual cortex \cite{Decoding2019}, which is rich in important visual information, and other multiple brain regions \cite{Decoding2024}, as well as overall studies on neural data collected across brain regions \cite{Multi-Scale2024}. 
However, this mode-separated research paradigm is still a limitedly explored field for understanding how neurons distributed in different brain regions represent natural scenes and the hierarchical structure and topological relationships of the brain itself. 

In this study, we utilized the Neuropixel dataset from the Allen Institute for Brain Science \cite{Siegle2021}, which contains the spike responses of hundreds of neurons from the mouse visual cortex and some subcortical brain regions. The main visual tasks were to decode the corresponding external visual stimuli, such as natural scenes and static gratings, from the neural data. Decoding neural data from a single brain region was defined as fine-grained decoding tests, while decoding neural data across brain regions was defined as rough-grained decoding tests. 
In the Fig. \ref{fig: intro}, We proposed an adaptive topological decoding method that uses deep network technology to explore the topological relationships of hierarchical visual data. The main contributions are as follows:

\begin{enumerate}
    \item We conducted a hierarchical classification of brain regions based on the amount of visual information extracted during fine-grained decoding tests, which is related to the information transmission capacity within these regions. Subsequently, we proposed an adaptive topological method (AT-ViT) that contains the hierarchical structure for decoding this cross-regional neural data. 
    \item We also found that the neural data collected from the hippocampus may even negatively impact performance enhancement. This finding is different from other studies that fuse data from various brain regions, and the specific role of hippocampal data with random decoding performance still holds significant scientific value.
\end{enumerate}

Finally, we validated the effectiveness of the proposed method through expensive experiments. Furthermore, this study introduces a new avenue for discussion regarding the utilization of hierarchical deep networks as a tool to elucidate the computational principle of the mouse visual system. 
We also hope that this paper can provide a new perspective for the field of biological vision to simultaneously apply hierarchical model structures and hierarchical data structures, such as brain's visual structures and brain's visual data.

\section{Related Work}

Generally, brain's visual data is initially encoded in retinal ganglion cells and further processed through encoding and decoding processes in the LGN and visual cortex. Then, they are processed to extract more abstract and meaningful information for learning and memory \cite{Reconstruction2020,Decoding2024}, as well as temporal response delays when moving up the visual hierarchy, with higher-level regions typically showing slower response times \cite{Hierarchical2019,Hierarchical2022}. These observations suggest that the amount of visual information encoded or captured in brain structures may have a certain hierarchical structure, that is, following a hierarchical organization principle. 

In recent years, deep neural network (DNN) models have become valuable tools in neuroscience research. Deep learning-based methods have been developed to explore visual system datasets in the mouse brain \cite{How2019,Developing2019}. In the neuroscience research paradigm, brain topology studies in neural decoding mainly focus on static graph construction \cite{dynamic2022,Classification2023} and dynamic graph construction \cite{Learning2021,Dynamic2021}. Among them, static graph construction using predefined patterns does not explicitly model the dynamic interdependencies between different regions of interest (ROIs) and is difficult to generalize in an end-to-end manner to different downstream tasks \cite{Multi-Scale2024}. However, dynamic construction holds significant value in the study of brain functions with time-varying characteristics. Then, the Allen Brain Science Datasets with rich visual tasks are highly representative. Neural data collected through physiological neural probes have high spatiotemporal resolution and can be used to study single brain regions with rich visual information, such as the VISp brain region \cite{Decoding2019}, as well as the visual hierarchy across multiple brain regions (from nuclei and visual cortex to the hippocampus) \cite{Decoding2024}. However, these studies focus mainly on information processing in single brain regions and topological processing in multiple brain regions, seemingly neglecting collaborative information processing between single brain regions and multiple brain regions. 

In addition, in the field of neural decoding, multiple studies have confirmed that the visual cortex regions of mice may represent semantic features of learned visual categories \cite{Siegle2021,Deep2021}. In addition to visual coding, the hippocampus of rodents is believed to play a role in learning and memory similar to that of primates \cite{Hippocampal2017}. Regarding the function of the hippocampus, some researchers based on DNN models have suggested that hippocampal neurons contain less pixel information than those in the thalamus and visual cortex \cite{Decoding2024}, and that the hippocampus can encode more abstract information or concepts \cite{Plugging2019}. In our study, we also found that neural data collected from the hippocampus under the same conditions hurt the category decoding of visual tasks, consistent with the random baseline.

\section{Method}

\begin{figure*}
    \centering
    \includegraphics[width=\textwidth]{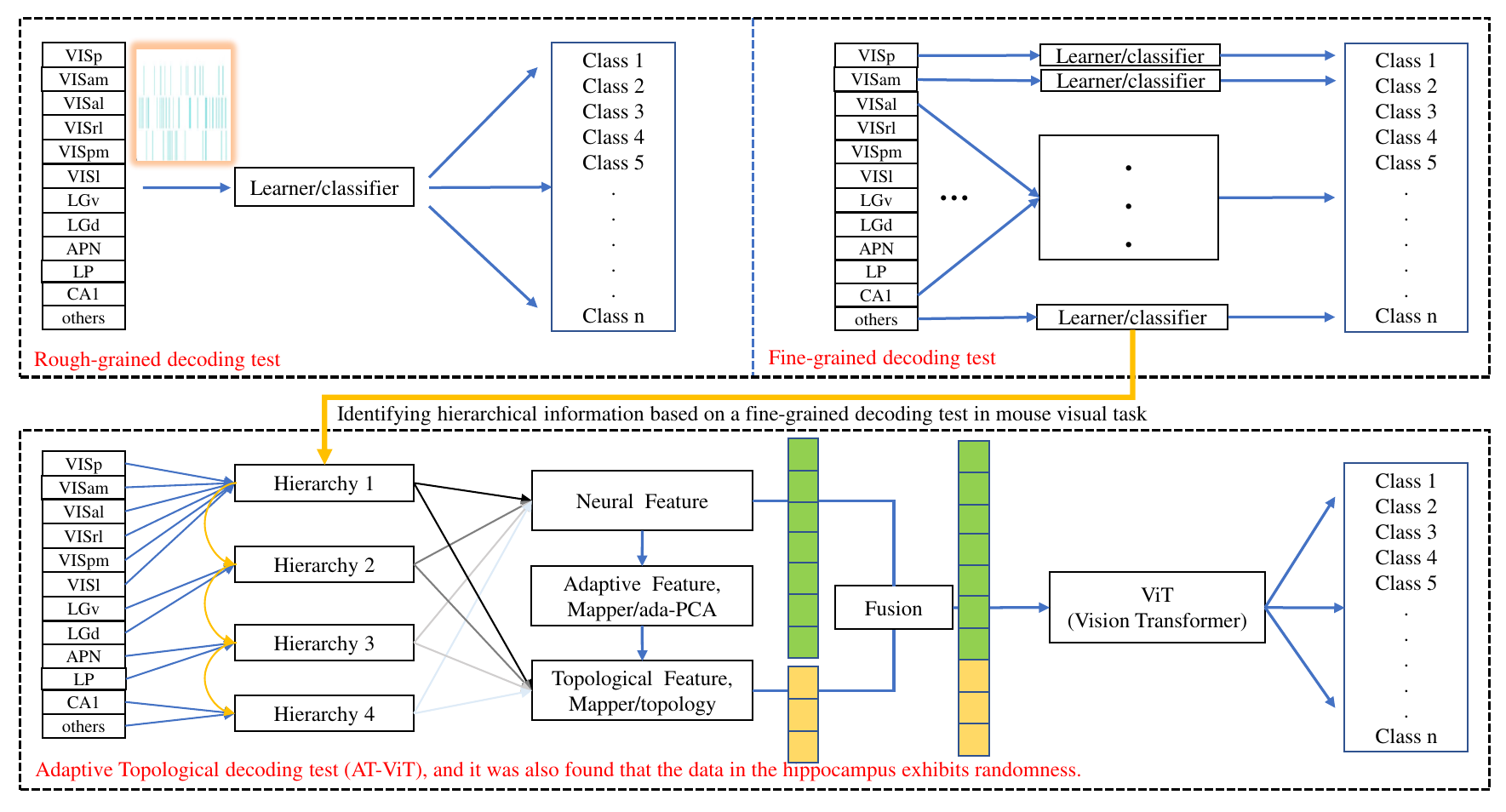}  
    \caption{Three methods are presented, namely rough-grained decoding tests, fine-grained decoding tests, and Adaptive Topological decoding. In rough-grained decoding tests, the focus is mainly on the topological analysis of brain's visual data. In fine-grained decoding tests, the emphasis is on the detailed information mining of each brain region. Adaptive Topological decoding essentially incorporates the ideas of the first two modes. That is, since the hierarchical brain's visual structure generates hierarchical brain's visual data, the important information extracted in fine-grained decoding tests can usually reveal this hierarchical structure. Based on this, the Adaptive Topological decoding model is proposed to handle hierarchical visual data in the brain's visual tasks.}
    \label{fig: framework}
\end{figure*}

\begin{figure}
    \centering
    \includegraphics[width=0.5\textwidth]{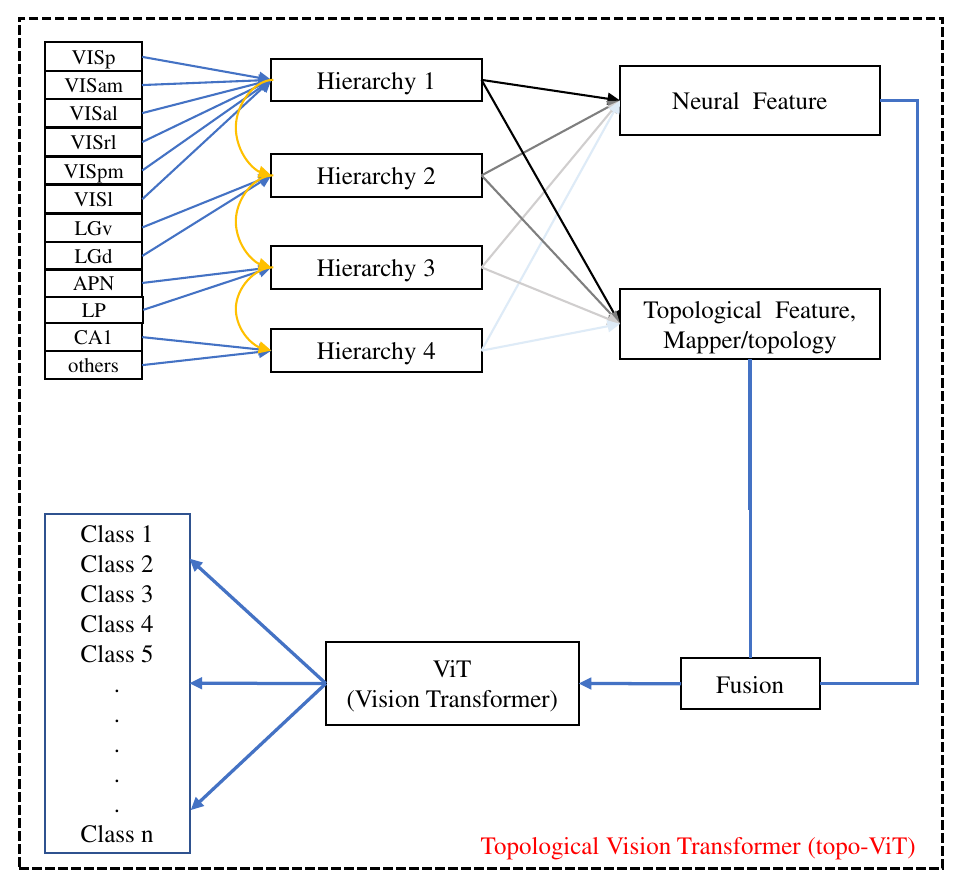}  
    \caption{Compared with AT-ViT, topo-ViT can be used for ablation experiments; it can be also used to verify whether there are factors such as noise in the neural data that can be eliminated to better extract topological features. Here, the ViT consists of 6 transformer layers, each with 32 attention heads and a hidden size of 64. In the feedforward network of the Transformer, the hidden layer is set to 4.
    }
    \label{fig: fra}
\end{figure}

\subsection{Pre-knowledge}

\subsubsection{Mapper Algorithm}

The Mapper algorithm \cite{Gurjeet2007,Hendrik2019} is a core tool in topological data analysis (TDA), mainly serving as an integrated approach for dimensionality reduction and clustering of data. Its advantage lies in preserving the topological features of the data and being capable of constructing and visualizing the topological structure of high-dimensional data. 

The basic steps behind Mapper are as follows \cite{Gurjeet2007}:
(1) Map to a lower-dimensional space using a filter function $F$, such as uniform manifold approximation and projection (UMAP)\cite{Leland2020}, in a given brain's visual data $D$.
(2) Construct a cover $(U_i)_{i\in I}$. $i$ is the $i$th interval in the total interval $I$.
(3) For each interval $U_i$ cluster the points in the preimage $F^{-1}(U_i)$ into sets $C_{i,1},...,C_{i,k_i}$. $k_i$ is the $k$th cluster set in the $i$th interval.
(4) Construct the graph whose vertices are the cluster sets and an edge exists between two vertices if two clusters share some points in common, which represents the topological structure of the data.

\subsubsection{Maximum Likelihood Estimate for PCA (ada-PCA)} 

Principal Component Analysis (PCA) is a commonly used data dimensionality reduction technique. 
Its advantages lie in the fact that by retaining the main components, it can filter out noise in the data for data preprocessing. 
To overcome the drawback of traditional PCA that requires manual specification of the number of principal components, an automatic method for selecting the dimension of PCA has been proposed \cite{Thomas2000}. This method, through Bayesian model selection \cite{Robert1995,MacKay1995}, can automatically select the most appropriate number of principal components or the optimal parameters based on the consideration of model complexity and the amount of data. 

We adopted the method proposed in \cite{Thomas2000}, which we refer to as ada-PCA here. Through Bayesian model selection techniques, the $n$-dimensional sample set $D = (\mathbf{s}_1, \mathbf{s}_2,..., \mathbf{s}_n)$ is automatically reduced to an $m$-dimensional sample set $D = (\mathbf{s}_1, \mathbf{s}_2,..., \mathbf{s}_m)$. $\mathbf{s}_k$ refers to the neural data collected by the $k$th moment in the brain's visual data. 

\subsubsection{Random Baseline in Mouse Visual Classification} 

A defined random baseline (Rb), or random accuracy, is mainly used to evaluate whether the model is above the correct prediction in a classification. Here, each class-label is assumed to be independent. The formula is as follows: 
\begin{equation}
Rb = \frac{1}{n},
\end{equation}%
where $Rb$ is an abbreviation for random baseline and $n$ refers to the number of classification labels in the classification task.

\subsection{Adaptive Topology Vision Transformer (AT-ViT)}

\subsubsection{Rough-grained Decoding Tests}

The rough-grained decoding described in Fig. \ref{fig: framework} refers to the process where the neural data collected from each brain region are sent together to the constructed model for processing. This approach is often used when it is necessary to explore the hierarchical relationships or topological structures among brain's visual system, as each brain region contains different amounts of information, resulting in a hierarchical relationship. As shown in the figure, the neural data collected from the visual cortex (VISp, VISam, VISal, VISrl, VISpm, VISl), thalamus/midbrain (LGv, LGd, APN, LP), and hippocampus (CA1, CA3, DG, SUB) are the firing data of the brain under visual stimulation, which have a one-to-one correspondence with the visual stimuli. Therefore, if all the neural data are sent to the model together, the trained model will be able to extract more visual information from the data to decode the visual stimuli. However, this method may overlook the generation process and discharge mechanism of the visual system. 

\subsubsection{Fine-grained Decoding Tests}

The fine-grained decoding described in Fig. \ref{fig: framework} refers to the fact that the neural data collected from each brain region will be sent to the constructed model for processing, respectively. This method is often used to explore the amount of information contained in each brain region. Generally speaking, in visual decoding tasks, the brain region VISp in the visual cortex is the neural data we commonly use because it is the primary cortex of the visual system, and almost all the information transmitted to the visual cortex passes through the primary cortex. Due to the hierarchical organization of the brain's visual system, that is, each brain region may only extract or process part of the information from the visual stimuli. Therefore, in addition to the VISp, the study of neural data from other brain regions is of great value. 

To better assess the amount of visual information contained in each brain region, we decoded the neural data collected from each brain region. These results are of significant reference value for us to extract visual information from individual brain regions. Additionally, if visual system indeed has a hierarchical organizational, then the results of visual decoding from individual brain regions should also exhibit a hierarchical organizational structure. This is a hypothesis.

\subsubsection{Adaptive Topological Decoding}

In Fig. \ref{fig: framework}, the adaptive topological decoding test (AT-ViT) describes that the neural data collected from each brain region will be assigned to a hierarchical structure and then sent to the constructed model for processing. Here, this assigned hierarchical structure is mainly based on the amount of visual information observed in fine-grained decoding tests. Each hierarchy contains the neural data of several brain regions. It can be seen from the figure or from the experimental results that hierarchy 1 mainly includes the brain regions of the visual cortex, hierarchy 2 includes the visual cortex and thalamus/midbrain (LGv, LGd), hierarchy 3 mainly includes the visual cortex and thalamus/midbrain, and hierarchy 4 includes the visual cortex, thalamus/midbrain, and hippocampus. In each hierarchy, the neural data will be adaptively dimensionally reduced and then the topological features will be extracted through the Mapper algorithm. These topological features and neural data are fused and sent to the Vision Transformer (ViT) model, attempting to absorb these hierarchical neural data through a hierarchical deep network.

In Fig. \ref{fig: fra}, the topological vision transformer (topo-ViT) describes that the neural data collected from each brain region will be assigned to a hierarchical structure and then sent to the constructed model for processing. Here, there is no adaptive dimensionality reduction processing. We attempt to directly extract topological features from the neural data and observe the decoding effect of this hierarchical deep network.

\subsection{Algorithms and Listings}

\begin{algorithm}[tb]
    \caption{AT-ViT Algorithm}
    \label{alg:algorithm}
    \textbf{Input}: brain's visual data $D = (\mathbf{s}_1, \mathbf{s}_2, \dots, \mathbf{s}_n)$ \\
    \textbf{Parameter}:$Hierarchy = n$, $Transformer \; layers = 6$, $heads = 32$, $hidden \; size = 64$, $learning \; rate = 1e-3$, $feedforward \; layers= 4$, $epochs = 30$, $n=1,2,...,j$.\\
    \textbf{Output}: AT-ViT model $M$
    \begin{algorithmic}[1]
        \STATE Initialize $\theta \gets \text{model parameters}$, $\text{optimizer}$. $t \gets 0$.
        \STATE $D_n \gets \text{SVM}(D) \text{ in fine-grained decoding tests}$
        \STATE $D_{topo} \gets \text{\{Mapper: ada-PCA, topology\}}(D_n)$
        \STATE $D_{fusion} \gets \text{Function }F_{fusion}(D_n, D_{topo})$
        \WHILE{$t \leq epochs$}
            \WHILE{$\text{Traversal on } D_{fusion} \text{ is not completed}$}
                \STATE Sample batch $(s_1, s_2, \dots, s_b) \sim D$.
                \STATE $E \gets \text{PatchEmbed}(s_1, s_2, \dots, s_b) + \text{PosEmbed}$.
                \FOR{$l = 1$ \TO $Transformer \; layers$}
                    \STATE $E \gets \text{MultiHeadAttention}(E, heads)$.
                    \STATE $E \gets \text{FeedForward}(E, feedforward \; layer)$.
                \ENDFOR
                \STATE $\hat{L} \gets \text{Predict}(E)$.
                \STATE $\mathcal{L} \gets \text{CrossEntropy}(\hat{L}, L_{\text{true}})$.
                \STATE $\theta \gets \text{Update}(\theta, \nabla_\theta \mathcal{L}, learning \; rate)$.
            \ENDWHILE
            \STATE $t \gets t + 1$.
        \ENDWHILE
        \STATE \textbf{return} $M(\theta)$.
    \end{algorithmic}
\end{algorithm}

Before applying the AT-ViT, a hierarchical data structure needs to be extracted. In this paper, the collected neural data can be observed to have a hierarchical structure in fine-grained decoding tests based on the amount of visual information, that is, there exists a hierarchical structure in the brain's visual system. First, the neural data is processed through adaptive dimensionality reduction, and then the topological structure contained in the neural data is extracted through the Mapper algorithm. Finally, the neural data and topological feature are fused and fed into the hierarchical deep network. This hierarchical deep network is used to extract the hierarchical structure of the neural data.

\section{Experiment}

\subsection{Dataset and Metric}
\label{data}

The electrophysiological dataset we used is from Allen Brain Visual Coding \cite{Siegle2021}, and its dataset and preprocessing code are in https://allensdk.readthedocs.io/en/latest/visual\_coding\_neuropixels.html. This dataset consists of 32 experimental sessions, each containing three hours of total experimental data, and the same protocol was used across different mice. These datasets include functional traces of individual spikes at the single-neuron level, which were recorded during multiple repeated trials of various natural (bears, trees, cheetahs, etc.) and artificial (drifting gratings, oriented bars, etc.) visual stimuli presented to the mice. In this paper, we focus on two visual classification tasks: natural scenes and static gratings. For fine-grained decoding tests, seven sessions were used, such as (session\_id) 761418226, 763673393, 773418906, 791319847, 797828357, 798911424, 799864342. Then, for the training, testing, and comparison with other methods of AT-ViT, two new sessions were used, such as (session\_id) 760345702, 762602078. In this paper, all evaluation metrics adopt a unified classification accuracy, that is, 
\begin{equation}
    ACC = 1 - \frac{F_{count\_nonzero}(Y_{prediction} - Y_{label})}{len(Y_{prediction})},
\end{equation}%
where, $F_{count\_nonzero}$ is a function for counting the non-zero elements in a vector, $Y_{prediction}$ is the predicted value of sample $Y$, $Y_{label}$ is the true value of sample $Y$, and $len(Y_{prediction})$ is the total length of sample $Y$.
Finally, 10-fold cross validation was used on the dataset to evaluate the model and verify the hypothesis.

\subsection{ada-PCA/SVM Decoding in Visual Tasks}

\begin{figure}
    \centering
    \includegraphics[width=0.5\textwidth]{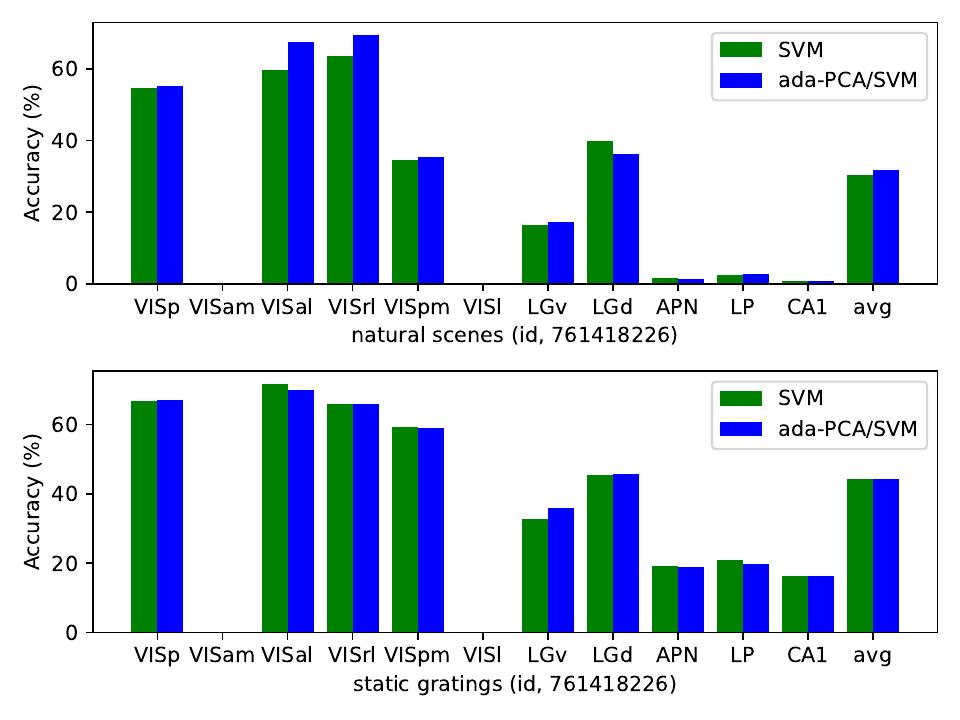}  
    \caption{Firstly, we employed a simple SVM algorithm and applied an adaptive dimensionality reduction algorithm (ada-PCA/SVM). The bar chart shows that the adaptive dimensionality reduction has certain advantages in brain's visual data.}
    \label{fig: nat/gra}
\end{figure}

Fig. \ref{fig: nat/gra} shows the decoding of brain's visual data using the adaptive dimensionality reduction and Support Vector Machine (SVM). In natural scenes, the histogram indicates that the decoding performance of ada-PCA/SVM, which has the ability of adaptive dimensionality reduction, is significantly higher than that of SVM alone. Additionally, the decoding performance among different brain regions also reveals that the amount of information contained in their data varies. Specifically, the decoding performance of the visual cortex is significantly higher than that of the thalamus/midbrain, and the thalamus/midbrain's performance is higher than that of the hippocampus. In static gratings, the decoding performance of ada-PCA/SVM is almost the same as that of SVM alone. In addition, the amount of information contained in each brain region shows a similar trend as in natural scenes.

\subsection{Fine-grained Decoding Tests in Visual Tasks}

\begin{table*}[htbp]
\caption{The visual (neural) data of mice are decoded by SVM in each brain area, and the reference random accuracy (\%) for natural scenes/static gratings is 0.78/16.67. Here, the random accuracy is used to evaluate the possibility of random prediction of the model, which has a significant reference value. For example, in natural scenes, the possibility of random prediction for 128 scene class-labels is $Rb = 1/128 = 0.78\%$; in static gratings, the possibility of random prediction for 6 direction class-labels is $Rb = 1/6 = 16.67\%$.}
\begin{center}
\begin{tabular}{|c|c|c|c|c|c|c|c|c|c|c|c|c|c|c|}
\hline
session\_id (nat) & VISp & VISam & VISal & VISrl & VISpm & VISl & LGv & LGd & APN & LP & CA1 & avg$\pm$std $\uparrow$\\
\hline
761418226 & \textbf{54.69} & -- & \textbf{59.85} & \textbf{63.71} & \textbf{34.62} & -- &  16.44 & 39.78 & 1.60 &  2.52 &  0.81 & 30.45$\pm$24  \\
763673393 & 46.67 & \textbf{32.03} & -- & 16.17 & -- & \textcolor{gray}{12.18} &  \textcolor{gray}{0.62} & \textbf{66.29} & \textbf{5.11} &  2.44 &  0.79 & 20.26$\pm$22  \\
773418906 & \textcolor{gray}{30.94} & 10.17 & 43.26 &  9.19 & --    & --    & --    &  --   &  3.18 &  --   &  0.76 & 16.25$\pm$16 \\
791319847 & 51.70 & 19.75 & 20.96 & 17.50 & 12.64 & 16.87 & 11.73 &  \textcolor{gray}{2.98} &    -- &  \textcolor{gray}{2.07} &  0.97 & 15.72$\pm$14 \\
797828357 & 32.17 &  \textcolor{gray}{7.12} &  \textcolor{gray}{3.72} &  \textcolor{gray}{5.60} &  \textcolor{gray}{6.50} & 14.67 & --    & --    &  3.38 &  8.07 &  0.74 & 9.11$\pm$09 \\
798911424 & 49.53 & 21.65 & 42.77 & 16.55 & -- & \textbf{39.56} & \textbf{37.45} & -- &  \textcolor{gray}{0.82} & \textbf{19.53} &  1.71 & 25.51$\pm$17 \\
799864342 & 50.29 & 27.48 & 24.66 & 12.08 & -- & 31.36 & -- & 51.08 &  0.94 & 14.62 &  0.76 & 23.70$\pm$18 \\
avg   & 45.14 & 19.70 & 32.54 & 20.11 & 17.92 & 22.93 & 16.56 & 40.03 &  2.51 &  8.21 &  0.93 & \textcolor{red}{20.60$\pm$14} \\
std       &  8.88 &  8.81 & 18.22 & 18.25 & 12.07 & 10.66 & 13.36 & 23.37 &  1.53 &  6.74 &  0.32 & 11.11$\pm$07\\
avg - ref(0.78)  & 44.36 & 18.92 & 31.76 & 19.33 & 17.14 & 22.15 & 15.78 & 39.25 & 1.73 & 7.43 & 0.15 & -- \\
\hline\hline
session\_id (gra) & VISp & VISam & VISal & VISrl & VISpm & VISl & LGv & LGd & APN & LP & CA1 & avg$\pm$std $\uparrow$\\
\hline
761418226 & 66.60 & -- & \textbf{71.72} & \textbf{66.00} & \textbf{59.19} & -- & 32.80 & 45.51 & 19.29 &  20.96 &  16.36 & 44.27$\pm$21 \\
763673393 & 60.88 & 52.02 & -- & 43.60 & -- & \textcolor{gray}{31.49} & \textcolor{gray}{14.80} & \textbf{60.67} & \textbf{25.99} & \textcolor{gray}{19.71} & 14.83 & 36.00$\pm$18 \\
773418906 & 54.69 & 36.89 & 63.27 & 32.72 & -- & --    & --    & -- & 20.31 & -- & 16.21 & 37.35$\pm$17 \\
791319847 & 65.15 & 51.99 & 42.23 & 59.29 & 38.44 & 41.76 & 31.72 & \textcolor{gray}{18.49} & -- & 20.72 & 16.82 & 38.66$\pm$16 \\
797828357 & \textcolor{gray}{48.98} & \textcolor{gray}{36.87} & \textcolor{gray}{23.35} & \textcolor{gray}{31.17} & \textcolor{gray}{25.59} & 31.99 & -- & -- & 22.95 & 21.50 & 16.85 & 28.81$\pm$09 \\
798911424 & 65.83 & \textbf{60.16} & 68.74 & 43.61 & -- & 58.40 & \textbf{47.17} & -- & \textcolor{gray}{16.66} & 34.33 & 20.29 & 46.13$\pm$18 \\
799864342 & \textbf{69.80} & 48.08 & 60.16 & 56.48 & -- & \textbf{58.86} & -- & 53.52 & 16.83 & \textbf{37.71} & 17.30 & 46.53$\pm$18 \\
avg   & 61.70 & 47.67 & 54.91 & 47.55 & 41.07 & 44.50 & 31.62 & 44.55 & 20.34 & 25.82 & 16.95 & \textcolor{red}{39.70$\pm$14} \\
std       &  6.87 &  8.43 & 16.97 & 12.41 & 13.84 & 12.11 & 11.47 & 15.97 &  3.31 &  7.30 &  1.54 & 10.02$\pm$05 \\
avg - ref(16.67)  & 45.03 & 31.00 & 38.24 & 30.88 & 24.40 & 27.83 & 14.95 & 27.88 & 3.67 & 9.15 & 0.28 & -- \\
\hline
\end{tabular}
\label{tab: nat/gra}
\end{center}
\end{table*}

Table \ref{tab: nat/gra} is based on Fig. \ref{fig: nat/gra}, presenting a simple SVM algorithm's performance in seven sessions and two visual tasks (i.e., natural scenes and static gratings), testing the amount of information contained in each brain region. The bold font indicates the highest decoding accuracy in each session and brain region. Moreover, to better evaluate the decoding performance and the amount of visual information in each brain region, a random decoding accuracy is used as a benchmark, i.e., the random baseline. If the decoding accuracy is close to the random baseline, the decoding of that brain region is equivalent to random guessing; if it is higher than the random baseline, the brain region contains more information. In natural scenes, it can be seen from the figure that the decoding accuracy of the visual cortex and thalamus/midbrain is significantly higher than the random baseline, while in the hippocampus, such as CA1, it is equivalent to the random baseline. This phenomenon is also reflected in the experimental results of static gratings. Similarly, in both natural scenes and static gratings, a hierarchical relationship regarding the amount of visual information can be observed based on the decoding performance, that is, the visual information in the visual cortex is higher than that in the thalamus/midbrain, and the visual information in the thalamus/midbrain is higher than that in the hippocampus.

\paragraph{Insight and discussion.} Based on the above analysis of the visual information contained in each brain region, the decoding performance of brain's visual data can reflect the hierarchical organization of the visual system. Based on this relationship, in fine-grained decoding tests, we can stratify brain's visual data based on the amount of visual information. Additionally, the random decoding accuracy observed in the hippocampus indicates that the visual information in this brain region is very scarce. However, if there is still visual discharge data in this brain region, can we assume that the neural data in this brain region might be a random guess?

\begin{table*}[htbp]
\caption{Based on the decoding results obtained in the brain's visual system, the amount of visual information contained in each brain area is evaluated and then divided into four hierarchical structures.}
\begin{center}
\begin{tabular}{|c|c|c|c|c|c|c|c|c|c|c|c|c|c|c|}
\hline
Hierarchy $n$ & VISp & VISam & VISal & VISrl & VISpm & VISl & LGv & LGd & APN & LP & CA1 & others (areas)\\
\hline
Hierarchy 1 & $\checkmark$ & $\checkmark$ & $\checkmark$ & $\checkmark$ & $\checkmark$ & $\checkmark$ & -- & -- & -- & -- & -- & --  \\
Hierarchy 2 & $\checkmark$ & $\checkmark$ & $\checkmark$ & $\checkmark$ & $\checkmark$ & $\checkmark$ & $\checkmark$ & $\checkmark$ & -- & -- & -- & --  \\
Hierarchy 3 & $\checkmark$ & $\checkmark$ & $\checkmark$ & $\checkmark$ & $\checkmark$ & $\checkmark$ & $\checkmark$ & $\checkmark$ & $\checkmark$ & $\checkmark$ & -- & --  \\
Hierarchy 4 & $\checkmark$ & $\checkmark$ & $\checkmark$ & $\checkmark$ & $\checkmark$ & $\checkmark$ & $\checkmark$ & $\checkmark$ & $\checkmark$ & $\checkmark$ & $\checkmark$ & $\checkmark$  \\
\hdashline
(id)760345702 & $\checkmark$ & $\checkmark$ & $\checkmark$ & -- & $\checkmark$ & $\checkmark$ & -- & $\checkmark$ & -- & $\checkmark$ & $\checkmark$ & $\checkmark$  \\
(id)762602078 & $\checkmark$ & $\checkmark$ & -- & $\checkmark$ & -- & -- & $\checkmark$ & -- & $\checkmark$ & $\checkmark$ & $\checkmark$ & $\checkmark$  \\
\hline
\end{tabular}
\label{tab: hierarchy/n}
\end{center}
\end{table*}

\begin{table*}[htbp]
\caption{In the four hierarchical structures that have been divided, the proposed model and other models are subjected to relevant experiments and comparative analysis in the two selected sessions.}
\begin{center}
\begin{tabular}{|c|c|c|c|c|c|c|c|c|c|c|c|c|c|c|}
\hline
session\_id (ada-PCA/SVM) & Hierarchy 1 & Hierarchy 2 & Hierarchy 3 & Hierarchy 4 & Mean (nat\_scenes/static\_gra) $\uparrow$ \\
\hline
760345702 & 87.83/83.19 & 90.84/84.77 & 91.18/84.65 & 89.41/82.24 & 89.82 (0\%) / 83.71 (0\%)  \\
762602078 & 95.31/85.21 & 95.92/85.42 & 95.46/84.11 & 93.83/82.06 & 95.13 (0\%) / 84.20 (0\%)  \\
\hdashline
session\_id (topo-ViT) \\
\hdashline
760345702 & 88.82/84.23 & 91.96/86.49 & 92.26/87.07 & 91.74/85.32 & 91.20 (1.54\%) / 85.78 (2.47\%) \\
762602078 & 95.63/87.43 & 96.23/86.84 & 96.18/86.98 & 95.71/86.53 & 95.94 (0.85\%) / 86.95 (3.27\%) \\
\hdashline
session\_id (AT-ViT)  \\
\hdashline 
760345702 & 89.10/84.51 & 92.48/86.04 & 92.60/86.57 & 92.01/85.97 & 91.55 (1.93\%) / 85.77 (2.46\%) \\
762602078 & 96.06/87.20 & 96.58/87.44 & 96.21/87.12 & 95.78/86.26 & 96.16 (1.08\%) / 87.01 (3.34\%) \\
\hline
\end{tabular}
\label{tab: Hierarchy/topology}
\end{center}
\end{table*}

\subsection{Brain Hierarchy Setting and Experiment}

\begin{figure}
    \centering
    \includegraphics[width=0.5\textwidth]{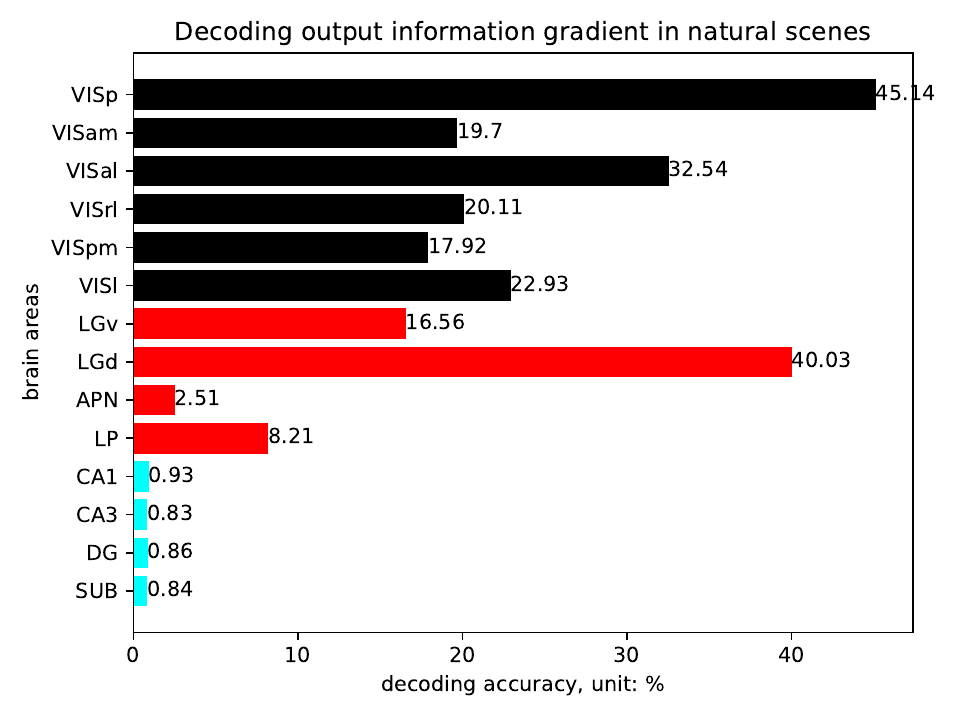}  
    \caption{The decoding output information gradient in natural scenes is presented. Among them, black represents the decoding accuracy of the visual cortex data, red represents the decoding accuracy of the thalamus/midbrain data, and cyan represents the decoding accuracy of the hippocampal data.}
    \label{fig: gradient}
\end{figure}

Tables \ref{tab: hierarchy/n} are based on Table \ref{tab: nat/gra}, and a hierarchical structure is divided based on the decoding effect of each brain region. As follows in Fig. \ref{fig: gradient}, \textbf{hierarchy 1}: visual cortex (VISp, VISam, VISal, VISrl, VISpm, VISl); \textbf{hierarchy 2}: visual cortex + thalamus/midbrain 1 (LGv, LGd); \textbf{hierarchy 3}: visual cortex + thalamus/midbrain 2 (LGv, LGd, APN, LP); \textbf{hierarchy 4}: visual cortex + thalamus/midbrain + hippocampus (CA1, CA3, DG, SUB). Here, some brain regions in the thalamus/midbrain are respectively classified into hierarchy 2 and hierarchy 3, because the visual information contained in the brain regions LGv and LGd sometimes aligns with that of the visual cortex, but the visual information in LGv, LGd, APN, and LP is significantly less than that in the visual cortex. 

Table \ref{tab: Hierarchy/topology} is based on the division of the hierarchy according to visual information based on Table \ref{tab: hierarchy/n}. The proposed method was trained, tested, compared and analyzed in two sessions. From the experimental results, the decoding performance of topo-ViT is higher than that of ada-PCA/SVM, with improvements of 1.54\%, 2.47\%, 0.85\%, and 3.27\% in natural scenes and static gratings, respectively. Additionally, the decoding performance of the proposed hierarchical networks (such as AT-ViT) is significantly higher than that of non-hierarchical methods (such as ada-PCA/SVM), with improvements of 1.93\%, 2.46\%, 1.08\%, and 3.34\% in natural scenes and static gratings, respectively. Overall, the performance of the proposed AT-ViT is also higher than that of topo-ViT. Finally, the decoding performance in hierarchy 1 and hierarchy 2 increases significantly, while in hierarchy 3, there is a slowdown trend, but in hierarchy 4, the performance significantly decreases. These results indicate that the hippocampus-related brain regions hurt the performance.

\subsection{Decoding and Analyzing in hierarchical information gradients}

\begin{figure}
    \centering
    \includegraphics[width=0.5\textwidth]{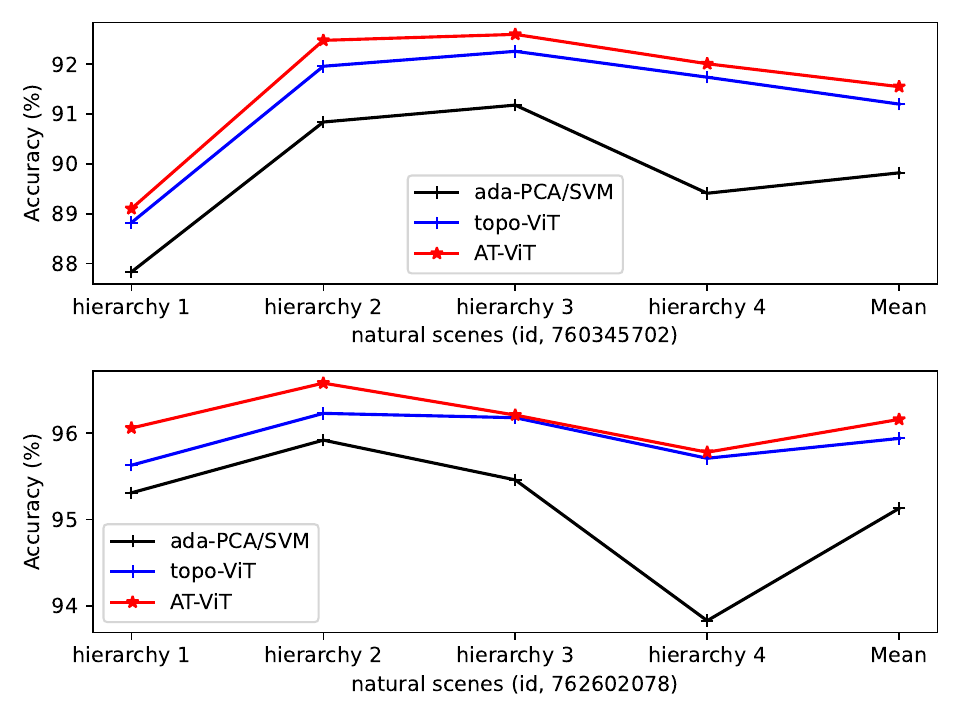}  
    \caption{Decoding accuracy map of two sessions in natural scenes
    }
    \label{fig: Hierarchy/topology}
\end{figure}

Fig. \ref{fig: Hierarchy/topology} shows the classification accuracy curves of visual tasks in the decoding of visual hierarchical data. Hierarchical networks significantly outperform non-hierarchical networks. In session (760345702), as the hierarchy increases, the decoding performance gradually improves, but it drops in hierarchy 4. This indicates that the neural data from the hippocampus (with performance close to the random baseline) hurts the decoding performance. In session (762602078), as the hierarchy increases, the decoding performance decreases in both hierarchies 2, 3 and 4. The reasons can be analyzed in Tables \ref{tab: nat/gra} and \ref{tab: hierarchy/n}. The performance in brain region LP is generally higher than that in APN, and even the performance in APN approaches the random baseline, such as 0.82\% and 16.66\%. It can be seen from Table \ref{tab: hierarchy/n} that session (762602078) collected neural data from APN and LP, and session (760345702) only record the neural data from LP. These experimental results reveal that the visual information contained in the visual data of different visual regions is highly complementary and holds significant reference value for joint research on multiple brain regions.
Based on these, the best decoding performance can be achieved when the hierarchical hyperparameter is set to $n = 3$ in this paper. That is to say, the neural data collected in the hippocampus during visual tasks is similar to noise, which affects the decoding performance of visual classification. In addition, we believe that the exploration and verification of other functions of the hippocampus can be further studied using other relevant datasets.

\section{Discussion}

This paper mainly explores brain's visual data from fine-grained decoding tests in a single brain region and rough-grained decoding tests across brain regions. The adaptive topological method (AT-ViT) is a preliminary attempt, and there is still much room for improvement in its model. For example, more advanced algorithms can be adopted for hierarchical data and combined with the hierarchical functions of the visual system. Currently, graph networks have certain advantages \cite{Multi-Scale2024}, but these networks do not well consider the characteristics of brain's visual system, that is, the generation process of hierarchical data. In the future, research on visual function can be conducted from the information attributes of visual data and the structural attributes of the visual system.

In the paper \cite{Decoding2024}, it was also found that the collected data from the hippocampus (including CA1, CA3, DG, and SUB) for the decoding of complex pixel-level details in the stimuli is unreliable. They believe that this difference can be partially attributed to their position at the end of the visual pathway. In our study, we defined a random baseline as a reference, and found that the collected data from the hippocampus may have a negative impact on performance, that is, random guessing. This hypothesis has scientific value for future research on the function of the hippocampus.

\section{Conclusion}

This paper mainly explores brain's visual data from the perspectives of fine-grained decoding tests in some single brain regions and rough-grained decoding tests across brain regions, and hierarchically divides the neural data based on visual information (or decoding outcomes). The proposed adaptive topological Vision Transformer (AT-ViT) initially addresses this hierarchical data and demonstrates the superiority of hierarchical networks in brain's visual data, mainly through adaptive dimensionality reduction and extraction of topological features to process visual data. Since brain's visual data originates from the hierarchical organization of the visual system, the amount of information contained in each brain region may vary, which has been proven in the experiment. In addition, this study also found that the neural data collected in the hippocampus may have a random baseline for decoding, which has a negative impact on decoding performance across brain regions. However, the specific function and firing mechanism of this hippocampal data with a random baseline still needs further research and have a scientific value.

\balance


\bibliographystyle{unsrt}












\newpage
\section*{Appendix A}
\newcommand{\red}[1]{\textcolor{red}{#1}}
\newcommand{\blue}[1]{\textcolor{blue}{#1}}

Some constructive suggestions are shown as follows for future research and improvement. 
Anonymous comments from others:

A1: This manuscript presents a compelling study on hierarchical neural decoding within the mouse visual system. The authors introduce an Adaptive Topological Vision Transformer (AT-ViT), which utilizes fine-grained decoding performance to construct a brain hierarchy, subsequently applying this hierarchy to cross-regional decoding. The key work (using decoding outcomes to infer hierarchical organization) is both novel and conceptually sound. The observation that hippocampal data impairs decoding performance to a level near the random baseline is particularly intriguing and invites meaningful discussion. Overall, the study addresses a significant problem in computational neuroscience through an innovative decoding framework, and the results (especially those pertaining to the hippocampus) are of considerable interest. I recommend minor revision to enhance the manuscript's clarity, methodological transparency, and interpretive rigor prior to publication. Specific Comments:

1. On page 2, left column, line 10, the phrase "expensive experiments" is unclear. Did the authors intend to say "extensive experiments"?

2. On page 2, left column, lines 23–27, the sentence beginning with "Then, they are processed…" is syntactically awkward and obscures the intended meaning. The connection between information abstraction, learning and memory, and temporal response delays is not logically articulated. This sentence should be revised for clarity and coherence.

3. The captions for Figure 4 and Figure 6 are incomplete and fail to adequately describe the content being presented. Please provide more detailed explanations to assist reader comprehension.

4. Clarity of Methodology:
a) The decoding task is not sufficiently detailed in the Methods. While the goal—decoding visual stimuli from neural data—is stated broadly, the specific formulation remains vague. The authors should explicitly define the input neural features (e.g., binned spike counts, firing rates over a specified time window), the output labels, the trial structure, and the temporal alignment between neural activity and stimulus presentation.
b) Although Algorithm 1 outlines the general AT-ViT and topo-ViT workflow, key procedural details remain unclear. Specifically, the processes of "adaptive feature" extraction and the "fusion" of neural and topological features require more thorough explanation to ensure reproducibility.

5. Interpretation of Hippocampal Findings: The claim that hippocampal data has a "negative impact" and is "similar to noise" may overinterpret the results. A more plausible explanation—consistent with existing literature—is that the hippocampus encodes information at a higher level of abstraction (e.g., contextual or mnemonic content) not directly relevant to the visual categorization task. Its activity may thus be uninformative in this context, rather than purely noisy. To strengthen this discussion, the authors should consider citing work on how high-level cognitive functions (e.g., working memory) modulate sensory areas. For example, the recent finding that visual working memory content is represented in primate V1 (Huang et al., Science Advances, 2024) suggests that top-down signals from higher-order regions can substantially influence early visual activity. \red{The seemingly random hippocampal signal in this task could reflect such task-irrelevant top-down modulation.}

6. Definition of the Hierarchy and Neurobiological Insight:
a) The rationale for assigning brain regions to specific hierarchies remains unclear. For instance, both the visual cortex and thalamic nuclei (e.g., LGv, LGd) are centrally involved in visual processing, yet they are placed in different hierarchies (e.g., Hierarchy 1 vs. 2). The authors should provide an explicit criterion—whether based on anatomical connectivity, functional specialization, or a quantitative decoding performance threshold—for these assignments.
b) The treatment of brain regions (especially the visual cortex) as monolithic units overlooks the rich functional substructure within individual areas. Acknowledging the existence of intra-areal hierarchical and specialized processing—such as the laminar-specific processing of spatial frequency (Wang et al., Nature Communications, 2024), orientation (Wang et al., Journal of Neuroscience, 2020), and luminance (Yang et al., Nature Communications, 2022)—would strengthen the biological plausibility of the proposed model. The authors are encouraged to discuss how their macro-scale, inter-areal hierarchy might relate to these well-established micro-architectures.

A2: The authors proposed an Adaptive Decoding via Hierarchical Neural Information Gradients in Mouse Visual Tasks framework and scheme, which has certain academic significance and application value.
However, the following concerns need to be addressed before the paper can be accepted for publication.

1) The introduction section lacks in-depth analysis of the challenges and difficulties of Mouse Visual Tasks, and many literature share the same analysis, which cannot constitute the motivation for this study.

2) In Figure 2, the visualization effect of the proposed technical framework is poor, lacking intuitiveness and logicality. The same issue exists in Figure 3. Suggest refining the drawing to highlight the visual effect.

3) In terms of ablation research, it is relatively comprehensive. However, the author has conducted relatively few experiments in comparison with other latest and mainstream methods, and it is suggested to supplement them. Regarding the experimental results, in addition to quantitative data tables, there is also a lack of visual displays such as ROC ,PR etc.

4) In the conclusion section of the paper, it is suggested to add a discussion on the limitations of this research work.

5) Authors are encouraged to open source algorithm code as much as possible.



\end{document}